# Orbital-Angular-Momentum-embedded Einstein–Podolsky–Rosen Entanglement Generated from Cold Atoms

Shuai Shi[1,2,†], Dong-sheng Ding[1,2,†,*], Yi-chen Yu[1,2], Wei Zhang[1,2], Ming-xin Dong[1,2], Kai Wang[1,2], Ying-hao Ye[1,2], Guang-can Guo[1,2], Bao-sen Shi[1,2,*]

[1]*Key Laboratory of Quantum Information, University of Science and Technology of China, Hefei, Anhui 230026, China*

[2]*Synergetic Innovation Center of Quantum Information and Quantum Physics, University of Science and Technology of China, Hefei, Anhui 230026, China*

[†]*These authors contributed equally to this work.*
[*]*Corresponding authors: Ding D-S dds@ustc.edu.cn; Shi B.S. drshi@ustc.edu.cn*

The Einstein–Podolsky–Rosen (EPR)-entangled quantum state is of special importance not only for fundamental research in quantum mechanics, but also for information processing in the field of quantum information. Previous EPR-entangled state demonstrations were constructed with photons of equal phase wave fronts. More complex scenarios with structured wave fronts have not been investigated. Here, we report the first experimental demonstration of EPR entanglement for photon pairs carrying orbital angular momentum (OAM) information, resulting in an OAM-embedded EPR-entangled state. We measured the dynamics of the dependence of the ghost interference on relative phase under projection. In addition, the reconstructed matrix in the OAM and EPR position–momentum spaces shows a specific hyper-entanglement in high dimension.

The famous Gedanken experiment proposed by Einstein, Podolsky and Rosen (EPR) was designed to prove that quantum theory is incomplete [1]. In brief, if two particles are perfectly correlated in position and anti-correlated in momentum, the pair exhibits an apparent paradox, in that the product of the conditional variances of positions and momenta violate the Heisenberg inequality $\Delta x \Delta p \geq \hbar/2$, indicating the failure of local realism. This thought experiment became known as the EPR paradox. In contrast to what EPR predicted, experiments based on various systems demonstrated non-local features of quantum mechanics, such as spin-entangled states [2], bulk crystals [3–5], squeezed field [6,7], atomic ensembles [8–10] etc.

Non-locality in quantum mechanics was demonstrated initially using spin-entangled states [2]. Later demonstrations of the EPR paradox exploited continuous variables, specifically, the quadrature-phase amplitudes of two-mode squeezed states [6,7]. The original EPR proposal with true position–momentum entanglement of photon pairs became available in spontaneous parametric down-conversion (SPDC) [4,5,11–13]. The reason why EPR entangled states are widely studied is that the EPR paradox plays a vital role in viewing the quantum world. A revolutionary prospective, based on quantum mechanical non-locality, has emerged with the development of quantum communication [14–16] and quantum computation [17].

Even though the EPR paradox in position and momentum degrees of freedom (DOFs) has been demonstrated, both of which are measured with equal-phase-wave-front photon pairs [3,12,13,16,18], there is no work reporting such measurements with structured phase wave-fronts. Furthermore, whether these special photon pairs affect the EPR position–momentum entanglement is unclear. In addition, photon pairs with helical phase structures have attracted much interest, because the OAM states of a photon enable encoding with inherent infinite DOFs, thereby enhancing the channel capacity and significantly improving the efficiency of a network [19-24]. Moreover, narrowband-photon pairs entangled in position–momentum DOF are promising for spatially-multiplexed quantum information processing, storage of quantum images, and quantum interface involving hyper-entangled photons [15,25–28]. Therefore, the fundamental question as to whether photons can be entangled in these two DOF is very interesting, and merits exploration.

Here we report the first experimental realization of the EPR position–momentum entanglement of photon pairs prescribed with vortex phase structures. The demonstration entails quantum ghost interference, ghost imaging, and the characterization of the OAM state via tomography. As there is a difficulty in measuring different DOFs simultaneously [23,29,30], we sort the OAM states using a Mach–Zehnder (M–Z) interferometer [31–33] and detect ghost interference via the OAM modes. The photon pairs for different vortex phase structures are shown to obey the EPR paradox inequality. We investigate the difference in ghost interferences between photon pairs with even and odd OAM that arises from a phase difference in the axisymmetric direction. We further demonstrate the hyper-entanglement in the OAM DOF and the linear momentum DOF. Photon pairs have been proved to be simultaneously entangled in multiple DOF-polarizations, spatial modes, and energy–time [34–38]. In this work, we demonstrate that linear momentum DOF can be added to the list. Hyper-entanglement in these two DOFs offers significant advantages in quantum communication protocols, e.g., secure superdense coding [39] and cryptography [40].

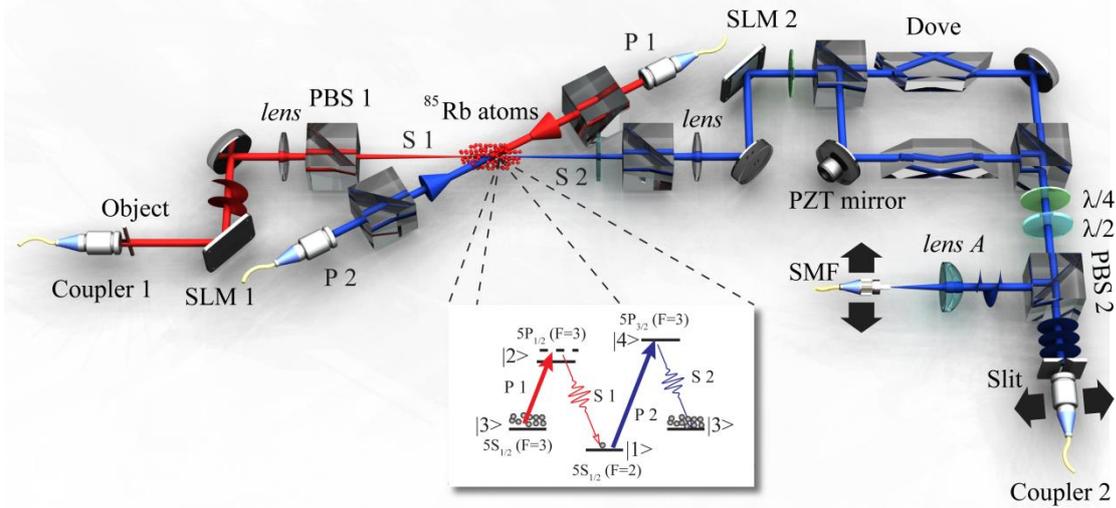

FIG. 1. Experimental setup: the photon pairs are generated by spontaneous four-wave mixing (SFWM) using counter-propagating pump fields 1 and 2. The SFWM is based on a double-Λ atomic configuration with energy levels of the ground state |1>, metastable state |3>, and excited states |2> and |4>, associated with $5S_{1/2}(F=2)$, $5S_{1/2}(F=3)$, $5P_{1/2}(F=3)$, and $5P_{3/2}(F=3)$ configurations, respectively; see inset for the energy diagram. The lenses ($f$=500 mm) are used to collimate the diverging Signals 1 and 2. The Signal-1 photon is first projected onto a spatial light modulator (SLM 1) and then imaged by a metal object, and finally collected by Coupler 1. The Signal-2 photon is reflected by SLM 2 and incident into an M-Z interferometer, which sorts photons of even and odd OAM into different paths for ghost interference and imaging.

With reference to the experimental setup and energy levels (Fig. 1), the photon pairs are prepared by continuous SFWM in a cold $^{85}$Rb atomic ensemble trapped in a two-dimensional magneto-optical trap (MOT). Laser pump 1 is blue-detuned by 65 MHz to the atomic transition |3> ($5S_{1/2}(F=3)$)→|2> ($5P_{1/2}(F'=3)$), and laser pump 2 is resonant with the atomic transition |1> ($5S_{1/2}(F=2)$)→|4> ($5P_{3/2}(F'=3)$). Their respective laser-beam power is 0.26 mW and 5 mW, respectively. The angle between the signal and pump laser beams is 2.5°. The polarizations of Signal 1 and Pump 2 are horizontal and vertical for Signal 2 and Pump 1; the setup is consistent with the SFWM matching condition of spin angular momentum. Our system operates periodically with a cycle time 10 ms, which consists of 8.5 ms for trapping and initial state preparing and 1.5 ms for generating entangled photon pairs. The temperature-controlled Fabry–Perot etalons (500-MHz full-width-at-half-maximum transmission bandwidth; 10-GHz free spectral range) are inserted into the optical paths of the signal 1/2 (75%/65% transmission) to reduce noise. Subsequently, the Signal-1 and Signal-2 photons were detected using two single-photon detectors (Avalanche diode, PerkinElmer SPCM-AQR-15-FC with ~50% efficiency). The outputs from both detectors were connected to a time-to-digital converter (Fast Comtec. P7888) with 1-ns bin-width to measure the cross-correlation function.

On the side of Signal 1, a metal object of width 1.04 mm is placed in front of Coupler 1, which makes a double-slit structured profile of Signal 1, and then is collected by a single mode fiber (SMF). On the side of Signal 2, the M–Z interferometer including two Dove prisms sort the even- and odd-OAM photons into different paths by rotating their polarizations perpendicularly to each other [33]. At PBS 2, transmission/reflection is selected for ghost imaging/ghost interference measurements. For ghost imaging, we scan a coupler at the image plane of the object. The resolution depends on the width of the slit in front, i.e., 0.4 mm. For ghost interference, we measure the correlation distribution at the focus plane of lens $A$ ($f$ = 25.4 mm) by scanning using a SMF tip. By these methods, we can obtain the uncertainties in position and momentum $\Delta x_-$ and $\Delta p_+$, thereby determining whether there is EPR entanglement between the Signal 1 and Signal 2 photons [13,41]. By inserting two SLMs in the two optical paths, we project the photons into OAM modes and measure the uncertainties, $\Delta x_-$ and $\Delta p_+$, and then verify EPR entanglement under different OAM modes.

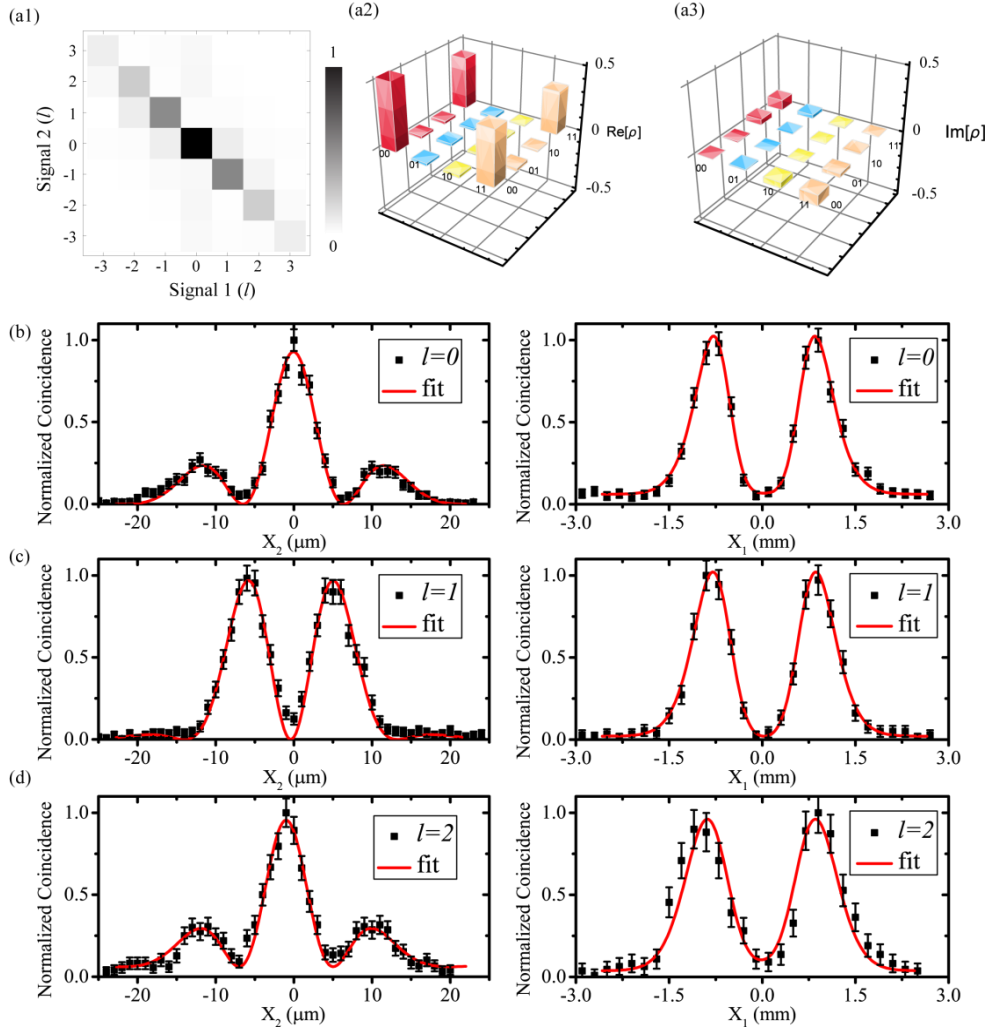

FIG 2. Demonstration of OAM and EPR entanglement: (a1) Measured distribution of correlated OAM modes ($l=-3–3$) from photon pairs. (a2,a3) are the density matrices corresponding to the OAM entangled state ($\psi_{01}=(|0\rangle|0\rangle+|1\rangle|1\rangle)/\sqrt{2}$) through tomography. (b, c) are the measured ghost interference and ghost images based on the OAM modes with $l=0, 1, 2$, respectively. Each point of the recorded data was accumulated in 30 seconds. The red curves given in these results are the simulated results from theory. The error bars represent statistical error of ±1 standard deviation.

As SFWM is a nonlinear process that conserves momentum, the initial state of the system has zero linear and zero angular momentum, so that the resulting joint state of the Signal 1 and Signal 2 photons has equal momentum that induces anti-correlations in the OAM and linear momentum states. Thus, the generated photon pairs should be entangled in the product space of OAM and linear momentum. The established entanglement is written

$$|\Psi\rangle = A\int d\vec{k}_{s1} d\vec{k}_{s2} C_\perp(\vec{k}_{s1},\vec{k}_{s2})|\vec{k}_{s1}\rangle|\vec{k}_{s2}\rangle \otimes (\sum_{l=-\infty}^{l=\infty} C_l |-l\rangle_{s1}|l\rangle_{s2}), \quad (1)$$

where $A$ is a normalization constant, $C_\perp(\vec{k}_{s1},\vec{k}_{s2})$ the correlation function in linear momentum, $\vec{k}_{si}$ the transverse component of the wave vector, which is proportional to transverse momentum ($\vec{p}_{si} = \hbar \vec{k}_{si}$), $|l\rangle_{si}$ denotes the OAM state of the *l* quantum eigenmode; subscripts *s1* and *s2* refer to the photons of Signal-1 and Signal-2, respectively, and $|C_l|^2$ is the excitation probability for different OAM modes (See Supplemental Material for details).

To investigate OAM entanglement and EPR entanglement individually, we first block one arm of the M–Z interferometer. For the OAM entanglement measurement, the metal object and the slit are removed, so that Signals 1 and 2 can be collected by Couplers 1 and 2 without blockage. The SLMs act as spatial post-selection tools for projecting Signal 1/2 onto different OAM modes. The measured correlated distribution between these OAM modes is obtained for the OAM modes *l* =−3–3; see Fig. 2(a1). To verify the quantum mechanical correlations under the OAM DOFs, for example *l*=0, 1, we project the Signal 1 and Signal 2 photons onto the basis vectors $|G\rangle$, $|L\rangle$, $(|G\rangle - i|L\rangle)/\sqrt{2}$, and $(|G\rangle + |L\rangle)/\sqrt{2}$ (here $|G\rangle$ and $|L\rangle$ are states corresponding to a well-defined OAM of 0 ℏ and 1 ℏ, respectively) using two SLMs. We obtain the corresponding 16 coincidence rates and then use them to reconstruct the density matrix. The real and imaginary parts of the reconstructed density matrix are shown in Fig. 2(a2,a3), respectively. We use formula $F = Tr(\sqrt{\sqrt{\rho_{exp}}\rho_{ideal}\sqrt{\rho_{exp}}})^2$ to calculate the fidelity of the reconstructed density matrix by comparing it with the ideal density matrix, which is of 89%. Here $\rho_{ideal}$ is the density matrix corresponding to the ideal OAM entangled state of $|\psi\rangle = (|G\rangle|G\rangle + |L\rangle|L\rangle)/\sqrt{2}$, and $\rho_{exp}$ is the reconstructed density matrix [42]. Second, we verify whether the photon pair is EPR-entangled in position–momentum by demonstrating ghost imaging and ghost interference. For the EPR entanglement measurement, the metal object and the slit are added, and the SLMs act as mirrors. Signal 1 is partially blocked by the metal object, and Signal 2 is measured by scanning the position distribution in one dimension. We hence obtain the profiles for both ghost imaging and ghost interference using the coincidence count rates for the Signal 1 and 2 photons; the normalized results are shown in Fig. 2 (b). The accumulated time for each point in the interference data is 30 s, and 200 s for imaging data. The high-contrast ghost interference and ghost imaging indicate a high-degree of EPR position–momentum entanglement. To verify that these photon pairs are indeed EPR-entangled, we need to check whether the EPR-paradox inequality is satisfied, that is, [13,41],

$$\langle(\Delta x_-)^2\rangle\langle(\Delta p_+)^2\rangle < \frac{1}{4}|\langle[x_{s1}, p_{s1}]\rangle|^2, \qquad (2)$$

where $\Delta x_- = x_{s1} - x_{s2}$ and $\Delta p_+ = p_{s1} + p_{s2}$, *x* and *p* represent the photon position and momentum, respectively, $\Delta$ represents the standard deviation. We obtain the uncertainties for position and momentum by fitting the data

[Fig 2(b)] with the theoretical functions for ghost interference and imaging; see Supplemental Material for details. From the interference data [Fig. 2(b)], we obtain $\Delta p_+ = 2.30 \pm 0.07 \hbar/mm$ and $\Delta x_- = 0.059 \pm 0.0019 mm$. We therefore have $(\Delta x_-)^2 (\Delta p_+)^2 = 0.0183 \pm 0.0017 \hbar^2 \ll \hbar^2$. Similarly, from the imaging data [Fig. 2(b)], we have $(\Delta x_-)^2 (\Delta p_+)^2 = 0.0016 \pm 0.0012 \hbar^2 \ll \hbar^2$. Both results satisfy the EPR-paradox inequality, which indeed demonstrates EPR entanglement between Signals 1 and 2.

TABLE I. Uncertainties for EPR entanglement under different OAM modes.

| OAM | | $\Delta p_+$ ($\hbar$/mm) | $\Delta x_-$ (mm) | $(\Delta x_-)^2 (\Delta p_+)^2$ ($\hbar^2$) |
|---|---|---|---|---|
| $l=1$ | Ghost imaging | 3.4±0.2 | 0.013±0.0056 | 0.0023±0.0016 |
| | Ghost interference | 2.8±0.1 | 0.057±0.001 | 0.0257±0.0017 |
| $l=2$ | Ghost imaging | 4.1±0.1 | 0.014±0.0046 | 0.0033±0.0023 |
| | Ghost interference | 2.2±0.18 | 0.045±0.003 | 0.010±0.003 |

Moreover, we measured ghost imaging and interference under different OAM modes (for example, $l$=1, 2). To do this, SLM 1 is used as a spatial post-selection tool instead of mirror, and SLM 2 as a mirror. As the OAM modes are correlated, and only photons with specific OAM can be collected by Coupler 1, we obtain ghost interference and ghost imaging by coincident measurements for different OAM modes; see Fig. 2(b, c). The ghost interference patterns are clearly different for photon pairs of even and odd OAM; the former has three peaks and the latter two peaks. The difference stems from the phase difference between the axially-symmetrical double slits for Signal 1 (to be explained below). The theoretical two-photon correlation functions for ghost interference and ghost imaging under even and odd OAM modes are given in the Supplemental Material. Fitting the experimental data with the theoretical functions produce the joint uncertainties $\Delta x_-$ and $\Delta p_+$ for OAM 1 and 2, respectively; see Table I. We find that all of the results satisfy the EPR-paradox inequality, indicating that EPR entanglement not only exists in the Gaussian mode, but also in other OAM modes.

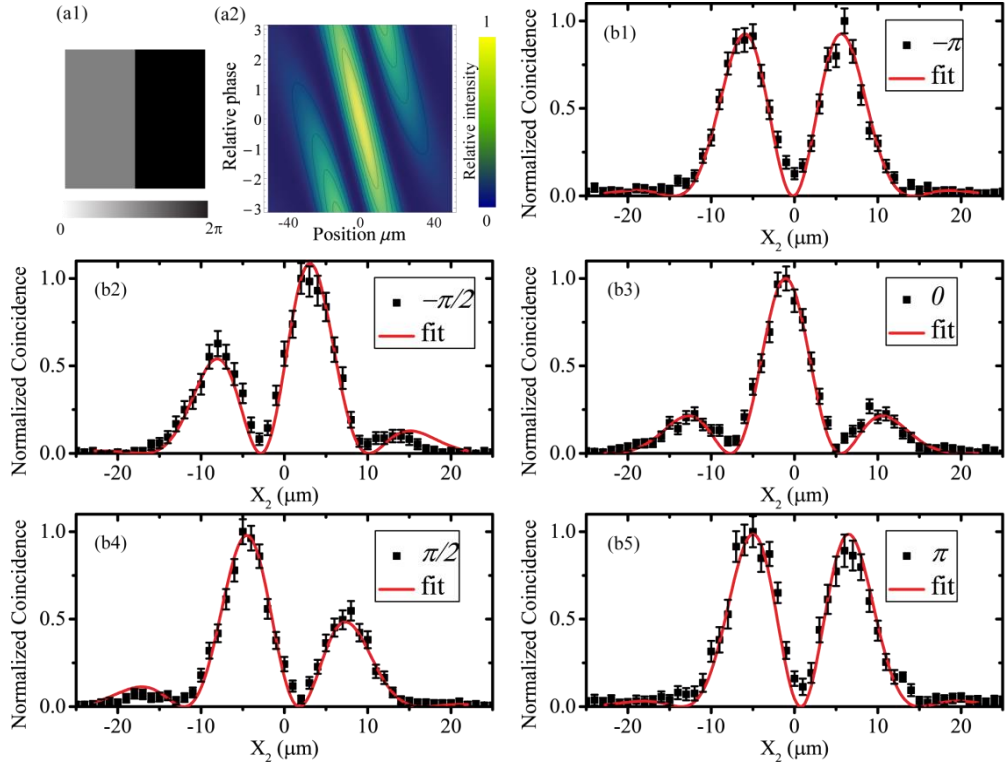

FIG. 3 Measured dynamics of EPR entanglement under different phase structure: (a1) phase structure loaded on the SLM 1 in the optical path of Signal 1; the relative phase $\theta$ is the phase difference between left and right parts of the phase structure. (a2) theoretical results of ghost interference under different relative phase $\theta$ from $-\pi$ to $\pi$. (b1–b5) measured data against different $\theta$; each point records data accumulated in 30 s. The red curves are theoretical fits; the error bars represent statistical error of ±1 standard deviation.

As the only obvious difference between photon pairs of even and odd OAM is that the phase in the central symmetric direction is the same or opposite, we conclude that the difference in their ghost interference profiles arises from this difference in relative phase. The axially-symmetrical double slits collect Signal-1 photons with the same or opposite phase in each slit window, which determines whether the entangled Signal 2 forms constructive or destructive interference at the symmetric center. To clarify how this phase difference affects the profiles, we measured the ghost interferences under different phase structures. The phase structure loaded onto SLM 1 shown in Fig. 3 (a1) was used to simulate the phase difference induced by OAM; the gray level represents the phase value. The phase difference $\theta$ between the double slits is compared with that between the left and right of the phase structure loaded on the SLM 1, which can be adjusted continuously. Taking the relative phase $\theta$ into consideration, the object transfer function for Signal 1 can be written

$$\Gamma(\vec{r}_1,\theta) = c_0 \exp(-\vec{r}_1^{\,2}/\omega_0^2)(1-\Pi(\vec{r}_1/\omega_b,\theta))\,, \qquad (3)$$

where $\omega_0$ is the Gaussian envelope, $\omega_b$ the width of the metal block, $\vec{r}_1$ the position vector on the Signal 1

detection plane, and $\Pi(x,\theta) = H(x+1/2) - \exp(i\theta)H(x-1/2)$ with $H(x)$ the Heaviside step function. By introducing a relative phase in the object transfer function, we can obtain a generalized two-photon correlation function for ghost interference; see Supplemental Material for details. Theoretical results for ghost interference under different relative phases $\theta$ from $-\pi$ to $\pi$ are shown in Fig. 3(a2).

TABLE II. Uncertainties for ghost interference under different relative phases.

| Relative phase | $\Delta p_+$ ($\hbar$/mm) | $\Delta x_-$ (mm) | $(\Delta x_-)^2(\Delta p_+)^2$ ($\hbar^2$) |
|---|---|---|---|
| $-\pi$ | 2.8±0.1 | 0.068±0.001 | 0.036±0.003 |
| $-\pi/2$ | 2.37±0.12 | 0.058±0.003 | 0.019±0.003 |
| $0$ | 2.9±0.1 | 0.059±0.003 | 0.030±0.004 |
| $\pi/2$ | 2.46±0.08 | 0.070±0.002 | 0.030±0.003 |
| $\pi$ | 2.8±0.1 | 0.067±0.002 | 0.036±0.004 |

We measured the ghost interferences for different relative phases ($-\pi$, $-\pi/2$, $0$, $\pi/2$, $\pi$) [Fig. 3(b1–b5)]. By fitting the experimental data with the theoretical function, we obtain the uncertainties for ghost interference under different relative phases (Table II). The EPR-paradox inequality is evidently satisfied for all phase structures demonstrating that despite the relative phase between the double slits varying, EPR entanglement remains. This supports the conclusion that the difference in ghost interference profiles for photon pairs with even and odd OAM indeed stems from the phase difference in the axisymmetric direction.

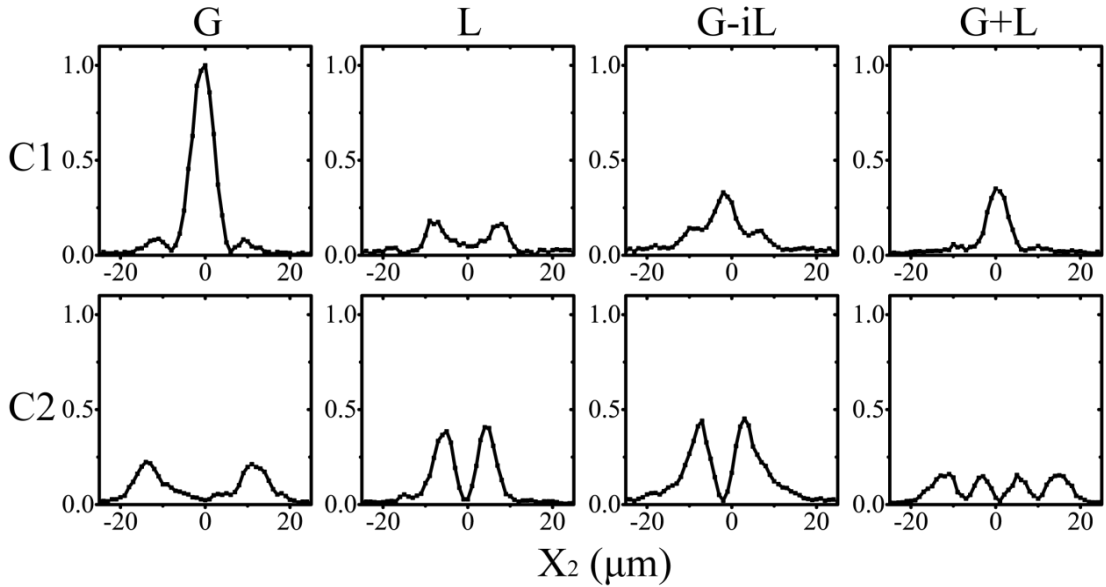

FIG. 4 Ghost interference data at the symmetric (C1) and anti-symmetric (C2) channel of the photon pairs with OAM states $|G\rangle$, $|L\rangle$, $(|G\rangle - i|L\rangle)/\sqrt{2}$, and $(|G\rangle + |L\rangle)/\sqrt{2}$.

Finally, we verify whether the OAM and EPR entanglement exist between the photon pair, simultaneously. In these experiments, the SLMs act as spatial post-selection tools for projecting Signal 1/2 onto different OAM modes, and the M–Z interferometer is fully integrated into the light path of Signal 2 rather than blocking one arm. We project the Signal-1 photons onto basis vectors $|G\rangle$, $|L\rangle$, $(|G\rangle - i|L\rangle)/\sqrt{2}$, and $(|G\rangle + |L\rangle)/\sqrt{2}$ using SLM 1. To avoid interference between the Gaussian mode and OAM $l$=1, we need to separate the two modes without any disruption before measuring the ghost interference. Here, we adopt the method based on M–Z interferometer to sort the OAM efficiently; the helical phase structure of OAM is also preserved [31–33]. We realize that state separation using an M-Z interferometer is based on its symmetry (see Supplemental Material for details). With the wave fronts of photons with even/odd OAM being center symmetric/anti-symmetric, respectively, the photons are routed to different channels (C1/C2). Note that the symmetry of the state is not only based on OAM DOF, but also related to the linear momentum DOF. If the state is anti-symmetric in the linear momentum DOF, $\frac{1}{\sqrt{2}}(E(\vec{k}) - E(-\vec{k}))$, even and odd OAM states will be routed to different channels. Hence the coincidence at the anti-symmetric channel (C2) for a Gaussian mode derives from the anti-symmetric state in the linear momentum DOF (see Fig. 4). If the linear momentum state is not anti-symmetric, a photon in a Gaussian mode will not be routed to channel C2; this was demonstrated using a coherent light field from Coupler 1 (see Supplemental Material for details). For the ghost interference data of the superposition states $(|G\rangle - i|L\rangle)/\sqrt{2}$ and $(|G\rangle + |L\rangle)/\sqrt{2}$, the anti-symmetric component of linear momentum interferes with the Gaussian component at C1, and interferes with the OAM $l$=1 component at C2. Hence, we have investigated exotic interference structures in OAM superposition (Fig. 4). The ghost interference on the superposition bases of Signal 2 were also measured (see Supplemental Material).

In summary, we demonstrated for the first time that EPR entanglement not only exists in the Gaussian mode, but also in other OAM modes. The ghost interference profiles for photon pair with even and odd OAM are different and stem from the different relative phase in axisymmetric direction. We also measured ghost interferences under different relative phases to study the dynamics of the EPR process; this can be further extended to photon pairs with non-integer charges because the relative phase in the axisymmetric direction is not an integer of $\pi$ for non-integer OAM. For photon pairs with special spatial structures formed as a superposition of different OAM modes, we can in principle implement EPR entanglement with arbitrary structured modes. Moreover, we can demonstrate hyper-entanglement in the OAM DOF and the position–momentum DOFs using tomography and the EPR-paradox inequality, respectively. The exotic interference structures arising from the superposition of OAM phases were investigated, and hold promise for a variety of quantum information protocols such as unconditional quantum teleportation [43,44], quantum key distribution [45] and high-dimensional communication [11,12,46]. Because the OAM and position–momentum DOFs correspond to infinite-dimensional Hilbert spaces—one is a discrete variable in the azimuth direction, and the other is a continuous variable in the linear momentum direction—the generated entanglement here is spanned in ultra-high-dimensional spaces. The estimation and limitation of such hyper-entangled dimensions are introduced in Supplemental Material. In addition, the reconstructed matrix in OAM and EPR position–momentum spaces indicates a specific hyper-entanglement spanned in the high dimension, which is significant in high-dimensional quantum communication, quantum computing, and quantum imaging.

This work was supported by National Key R&D Program of China, and funded by the National Natural Science Foundation of China ((Grant Nos. 61525504, 61722510, 61435011, 11174271, 61275115, 11604322).

# Supplementary Information of

# "Orbital Angular Momentum embedded Einstein-Podolsky-Rosen Entanglement Generated from Cold Atoms"

Shuai Shi[1,2,†], Dong-sheng Ding[1,2,†, §], Yi-chen Yu[1,2], Wei Zhang[1,2], Ming-xin Dong[1,2], Kai Wang[1,2], Ying-hao Ye[1,2], Guang-can Guo[1,2], Bao-sen Shi[1,2, *]

[1]Key Laboratory of Quantum Information, University of Science and Technology of China, Hefei, Anhui 230026, China

[2]Synergetic Innovation Center of Quantum Information and Quantum Physics, University of Science and Technology of China, Hefei, Anhui 230026, China

Correspondence: §Ding D-S dds@ustc.edu.cn; *Shi B.S. drshi@ustc.edu.cn

[†]These authors are contributed equally to this work.

**Estimation of the hyper-entangled dimension.** For OAM DOF, the photon pairs generated by SFWM in a cold atomic ensemble could entangled in a high-dimensional OAM, in principle, the number of dimensions can be simply estimated by the formula of $w(z) = \sqrt{m+1} w_0(z)$, where $w(z)$ is the beam waist of a light carrying OAM of $m$ at the center of the atomic ensemble, and $w_0(z)$ is the beam waist of Gaussian beam. In our experiment $m \sim 100$, therefore the maximal OAM dimension per photon that could generate in our system is limited to be 200. This number is also limited by other factors, such as the Fresnel number and the optical depth of the atomic ensemble, etc., which need further investigation. For EPR DOF, the dimension of entanglement $D = (\sigma_+ / \sigma_-)^2$, where $\sigma_+$ and $\sigma_-$ are the standard deviation of the correlation functions for momentum and position respectively [1]. According to the method in [2], in our experiment, the dimension of EPR entanglement is estimated as $D=54.6 \pm 0.5$. This number is limited by factors as the spatial profile of the pump beams and the angle of the signal field and the pump field, etc., which needs further investigation. Therefore, the photon pairs generated in our system can be in ultra-high dimensional by hyper-entangling two high DOFs, which is limited to be $1.08 \times 10^4$.

**Quantum state of photon pair**

The geometries shown in Fig. 1 is a backward propagation configuration for pump laser. The electric field is written as $E = \frac{1}{2}[E^{(+)} + E^{(-)}]$, where $E^{(+)}$ and $E^{(-)}$ are the positive and negative-frequency parts. The input strong classical field P1 and P2 are denoted as,

$$E_{p1}^{(+)}(z,t) = E_{p1}e^{i(-k_{p1}z-\omega_{p1}t)},$$

$$E_{P2}^{(+)}(z,t) = E_{p2}e^{i(k_{p2}z-\omega_{p2}t)}, \tag{S1}$$

Where $k_{p1,p2}$ are pump 1 and 2 field wavenumbers. The signal 1 and 2 field are quantized [3],

$$\hat{E}_{s1}^{(+)}(z,t) = \frac{1}{\sqrt{2\pi}}\int d\omega \sqrt{\frac{2\hbar\omega}{c\varepsilon_0 A}}\hat{a}_{s1}(\omega)e^{i[-k_{s1}(\omega)z-\omega t]}e^{-il_{s1}\alpha},$$

$$\hat{E}_{s2}^{(+)}(z,t) = \frac{1}{\sqrt{2\pi}}\int d\omega \sqrt{\frac{2\hbar\omega}{c\varepsilon_0 A}}\hat{a}_{s2}(\omega)e^{i[k_{s2}(\omega)z-\omega t]}e^{il_{s2}\alpha}, \tag{S2}$$

Where $k_{s1,s2}$ are wavenumber of signal 1 and 2 photons. $A$ is the single mode cross-section area. $\hat{a}_{s1,s2}$ are annihilation operators. $l_{s1,s2}$ are quantum number of orbital angular momentum, and $\alpha$ is the azimuthal angle. The effective interaction Hamiltonian for the SFWM process takes the form [4]

$$\hat{H}_I = \frac{\varepsilon_0 A}{4}\int_{-L/2}^{L/2}dz\int_{-\pi}^{\pi}d\alpha\chi^{(3)}E_{p1}^+E_{p2}^+E_{s1}^-E_{s2}^- + H.c., \tag{S3}$$

Where $H.c.$ stands for the Hermitian conjugate. $\chi^{(3)}$ is the third-order nonlinear susceptibility. Take the Eqs. (S1) and (S2) into (S3), and after the $z$ and $\alpha$ integration we rewrite the Hamiltonian as

$$\hat{H}_I = \frac{i\hbar L}{2}\int d\omega_{s1}d\omega_{s2}\kappa(\omega_{s1},\omega_{s2})\text{sinc}(\frac{\Delta kL}{2})\text{sinc}(\Delta l\pi)$$
$$\times \hat{a}_{s1}^\dagger(\omega_{s1})\hat{a}_{s2}^\dagger(\omega_{s2})e^{-i(\omega_{p1}+\omega_{p2}-\omega_{s1}-\omega_{s2})t} + H.c. \tag{S4}$$

Where $\Delta k = k_{s1} - k_{s2} - (k_{p1} - k_{p2})$ is the longitudinal phase mismatching. $\Delta l = l_{s1} - l_{s2}$ is the azimuthal phase mismatching. According to perturbation theory the two-photon state $|\Psi\rangle$ can be expressed as

$$|\Psi\rangle = -\frac{i}{\hbar}\int dt\hat{H}_I|0\rangle, \tag{S5}$$

Using the Hamiltonian (S4) in Eq. (6), the time integral yields a $\delta$ function, $2\pi\delta(\omega_{p1}+\omega_{p2}-\omega_{s1}-\omega_{s2})$, which expresses the energy conservation for the process. If $L$ is infinite, then the sinc function $L\text{sinc}(\frac{\Delta kL}{2})$ in (S4) becomes a $\delta$ function, $\delta(\Delta k)$, which expresses the linear momentum conservation. If OAM $l$ is integer, then the sinc function

$2\pi \text{sinc}(\Delta l \pi)$ in (S4) becomes a $\delta$ function, $\delta(\Delta l)$, which expresses the OAM conservation. In this case, the conditions

$$\omega_{p1} + \omega_{p2} - \omega_{s1} - \omega_{s2} = 0,$$

$$k_{s1} - k_{s2} - (k_{p1} - k_{p2}) = 0$$

$$l_{s1} - l_{s2} = 0 \tag{S6}$$

all hold and the phase matching is perfect. Now the two-photon state becomes [4]

$$|\Psi\rangle = \sum_{l=-\infty}^{l=\infty} C_l 2\pi L \int d\omega_{s1} \kappa(\omega_{s1}, \omega_{p1} + \omega_{p2} - \omega_{s1}) \text{sinc}(\frac{\Delta k L}{2}) \text{sinc}(\Delta l \pi)$$
$$\times \hat{a}_{s1}^\dagger(\omega_{s1}) \hat{a}_{s2}^\dagger(\omega_{p1} + \omega_{p2} - \omega_{s1}) |0\rangle \tag{S7}$$

From Eq. (S7) we can see that the two-photon state is entangled in frequency, wavenumber and OAM. According to Eq. (S7) and Ref.[1] the quantum state generated from SFWM can be written as

$$|\Psi\rangle = \sum_{l=-\infty}^{l=\infty} C_l A \int d\omega_{s2} d\vec{k}_{s1} d\vec{k}_{s2} \chi^{(3)}(\omega_{s2}, \omega_{s1}) \text{sinc}(\Delta k L / 2) \text{sinc}(\Delta l \pi)$$
$$\times \varsigma_\perp(\vec{k}_{s1}, \vec{k}_{s2}) \exp(-i(l_{s1} - l_{s2})\alpha) \hat{a}_{s1}^\dagger \hat{a}_{s2}^\dagger |0\rangle \tag{S8}$$

where $|C_l|^2$ is the excitation probability for different OAM modes, and

$$\varsigma_\perp(\vec{k}_{s1}, \vec{k}_{s2}) = \varepsilon_p(|\vec{k}_1 + \vec{k}_2|) \varepsilon_x(|\vec{k}_1 - \vec{k}_2|/2),$$

here $\varepsilon_p(|\vec{k}_1 + \vec{k}_2|) = \frac{1}{\sqrt[4]{\pi\sigma_p^2}} \exp(-\frac{|\vec{k}_1 + \vec{k}_2|^2}{2\sigma_p^2})$, $\varepsilon_x(|\vec{k}_1 - \vec{k}_2|/2) = \frac{1}{\sqrt[4]{\pi\sigma_x^2}} \exp(-\frac{(|\vec{k}_1 - \vec{k}_2|/2)^2}{2\sigma_x^2})$; $\varepsilon_p$, $\varepsilon_x$ are Gaussian functions for the correlation in momentum basis. Where $\sigma_p$ and $\sigma_x$ are the standard deviations of $\varepsilon_p$, $\varepsilon_x$, respectively.

$$\Delta(\vec{p}_{s1} + \vec{p}_{s2}) = \frac{\hbar\sigma_p}{\sqrt{2}}, \quad \Delta(\vec{x}_{s1} - \vec{x}_{s2}) = \frac{1}{\sqrt{2}\sigma_x} \tag{S9}$$

the quantum state can be simply written as

$$|\Psi\rangle = A \int d\vec{k}_{s1} d\vec{k}_{s2} C_\perp(\vec{k}_{s1}, \vec{k}_{s2}) |\vec{k}_{s1}\rangle |\vec{k}_{s2}\rangle \otimes (\sum_{l=-\infty}^{l=\infty} C_l |-l\rangle_{s1} |l\rangle_{s2}) \tag{S10}$$

Where $C_\perp(\vec{k}_{s1}, \vec{k}_{s2}) = \varsigma_\perp(\vec{k}_{s1}, \vec{k}_{s2}) \chi^{(3)}(\omega_{s1}, \omega_{s2}) \text{sinc}(\Delta k L / 2)$ is the correlation function in linear momentum.

**Calculation for ghost interference**

On Signal 1 side, the metal object imposes a constraint in transverse space at the detection plane, which can be represented as transmittance function $\Gamma(\vec{r},\theta)$, See Eq.(3) in the paper. The field at the detection plane of signal 1 can be calculated as [1],

$$E_{s1}^{(+)}(\vec{x}_{s1},z_{s1},t_{s1}) = \sum_{l=-\infty}^{l=\infty} C_{l,s1} \int d\omega d\vec{k}_{s1} \exp(-i\omega(t_{s1}-z_{s1}/c)) G(|\vec{x}_{s1}|)_{[\omega/cf_{s1}]} \\ \times \Gamma(\frac{\lambda f}{2\pi}\vec{k}_{s1},\theta) \exp(-i\frac{f}{f_{s1}}\vec{k}_{s1}\cdot\vec{x}_{s1}) \exp(il_{s1}\alpha) a_{s1} \quad (S11)$$

where $a_{s1}|\vec{k}_{s1}\rangle|l_{s1}\rangle = |0\rangle$. Since that the detector for Signal 1 is located at $r=0$, only photon with equal phase wave-front could be detected after the metal object. The transmittance function with different phase can be expressed as, $\Gamma(\vec{r},\theta) = \sum_{l=-\infty}^{l=\infty} c_l \Gamma(\vec{r},0) \exp(il\alpha)$. Due to the nonzero relative phase $\theta$ in the transmittance function, Signal 1 with different OAM would be detected. On signal 2 side, it is required to calculate the optical transfer function including two lenses, $l$ and $l_A$. The field at the detection plane of signal 2 can be written as,

$$E_{s2}^{(+)}(\vec{x}_{s2},z_{s2},t_{s2}) = \sum_{l=-\infty}^{l=\infty} C_{l,s2} \int d\omega d\vec{k}_{s2} \exp(-i\omega(t_{s2}-z_{s2}/c)) G(|\vec{x}_{s2}|)_{[\omega/cf_A]} \\ \times \exp(-i\frac{f}{f_A}\vec{k}_{s2}\cdot\vec{x}_{s2}) \exp(il_{s2}\alpha) a_{s2} \quad (S12)$$

The ghost interference is studied by the coincidence of the detection events of two single-photon detectors. The coincidence count rate reflects the second-order correlation. The joint probability can be written as [5],

$$G^{(2)}(\vec{x}_{s1},t_1,\vec{x}_{s2},t_2) = \left|\langle 0|\vec{E}_{s1}^{(+)}(\vec{x}_{s1},t_1)\vec{E}_{s2}^{(+)}(\vec{x}_{s2},t_2)|\Psi\rangle\right|^2 \quad (S13)$$

We can calculate the two-photon correlation function by inserting Eq. (S8), Eq. (S11) and Eq. (S12) into Eq. (S13), and assuming that the signal 1 is located at $x_{s1}=0$. Due to the entanglement in OAM DOF, the relative phase in the transmittance function of Signal 1 will arise in Signal 2. The correlation function for ghost interference can be calculated to be [1],

$$G^{(2)}(x_{s2}) \propto \left| \frac{\sigma_p \sigma_x \omega_0}{\sqrt{8\pi^2 \omega_0^2 + f^2(\sigma_p^2 + 4\sigma_x^2)\lambda^2}} \exp\left[ -\frac{f^2 x_{s2}^2 (\pi^2(\sigma_p^2 + 4\sigma_x^2)\omega_0^2 + 2f^2\sigma_p^2\sigma_x^2\lambda^2)}{f_A^2(8\pi^2\omega_0^2 + f^2(\sigma_p^2 + 4\sigma_x^2)\lambda^2)} \right] \right.$$

$$\times \left( \mathrm{erfc}\left[ \frac{-2if^2\pi(\sigma_p^2 - 4\sigma_x^2)\omega_0^2 \lambda x_{s2} + f_A\omega_b(8\pi^2\omega_0^2 + f^2(\sigma_p^2 + 4\sigma_x^2)\lambda^2)}{2ff_A\omega_0\lambda\sqrt{(\sigma_p^2 + 4\sigma_x^2)(8\pi^2\omega_0^2 + f^2(\sigma_p^2 + 4\sigma_x^2)\lambda^2)}} \right] \right.$$

$$\left. \left. + e^{i\theta} \mathrm{erfc}\left[ \frac{2if^2\pi(\sigma_p^2 - 4\sigma_x^2)\omega_0^2 \lambda x_{s2} + f_A\omega_b(8\pi^2\omega_0^2 + f^2(\sigma_p^2 + 4\sigma_x^2)\lambda^2)}{2ff_A\omega_0\lambda\sqrt{(\sigma_p^2 + 4\sigma_x^2)(8\pi^2\omega_0^2 + f^2(\sigma_p^2 + 4\sigma_x^2)\lambda^2)}} \right] \right) \right|^2 \quad \text{(S14)}$$

where $\theta$ is the phase difference between the double slits. For photon pairs with even and odd OAM, $\theta$ is 0 and $\pi$ respectively. By fitting the experimental data to Eq. (S14), we can obtain $\sigma_p$ and $\sigma_x$. Following Eq. (S9), the standard deviation of the total momentum $\Delta(\vec{p}_{s1} + \vec{p}_{s2})$ and position $\Delta(\vec{x}_{s1} - \vec{x}_{s2})$ can be calculated.

**Calculation for ghost imaging**

Now we consider the two-photon correlation function for ghost imaging. Signal 1 side is the same as that of ghost interference; the quantized field operator at the detection plane is the same as Eq. (S11). On Signal 2 side, the detection of ghost interference has been modified to scan the Fourier plane of the source. The quantized field operator at the imaging plane can be calculated as,

$$E_{s2}^{(+)}(\vec{x}_{s2}, z_{s2}, t_{s2}) = \sum_{l=-\infty}^{l=\infty} C_{l,s2} \int d\omega d\vec{k}_{s2} \exp(-i\omega(t_{s2} - z_{s2}/c)) G(|\vec{x}_{s2}|)_{[\omega/cf]}$$
$$\exp(il_{s2}\alpha) a_{s2} G(|\vec{k}_{s2}|)_{[-cf/\omega]} \delta(\vec{k}_{s2} - \frac{\omega}{cf}\vec{x}_{s2}) \quad \text{(S15)}$$

The two-photon amplitude can be calculated by using Eq. (S8), (S11), (S13) and (S15). The calculation result in,

$$G^{(2)}(x_{s2}) \propto \left| \frac{\sigma_p \sigma_x \omega_0}{\sqrt{2\pi^2(\sigma_p^2 + 4\sigma_x^2)\omega_0^2 + 4f^2\sigma_p^2\sigma_x^2\lambda^2}} \exp\left[ -\frac{2\pi^2 x_{s2}^2(8\pi^2\omega_0^2 + f^2(\sigma_p^2 + 4\sigma_x^2)\lambda^2)}{f^2\lambda^2(2\pi^2(\sigma_p^2 + 4\sigma_x^2)\omega_0^2 + 4f^2\sigma_p^2\sigma_x^2\lambda^2)} \right] \right.$$

$$\times \left( \left( 1 - \mathrm{erf}\left[ \frac{2f_a\pi^2\omega_0^2(4\sigma_x^2(\omega_b - 2x_{s2}) + \sigma_p^2(\omega_b + 2x_{s2})) + 4f^2 f_a \sigma_p^2\sigma_x^2\omega_b\lambda^2}{4ff_a\sigma_p\sigma_x\omega_0\lambda\sqrt{2\pi^2(\sigma_p^2 + 4\sigma_x^2)\omega_0^2 + 4f^2\sigma_p^2\sigma_x^2\lambda^2}} \right] \right) \right.$$

$$\left. \left. + e^{i\theta}\left( 1 - \mathrm{erf}\left[ \frac{2f_a\pi^2\omega_0^2(4\sigma_x^2(\omega_b + 2x_{s2}) + \sigma_p^2(\omega_b - 2x_{s2})) + 4f^2 f_a \sigma_p^2\sigma_x^2\omega_b\lambda^2}{4ff_a\sigma_p\sigma_x\omega_0\lambda\sqrt{2\pi^2(\sigma_p^2 + 4\sigma_x^2)\omega_0^2 + 4f^2\sigma_p^2\sigma_x^2\lambda^2}} \right] \right) \right) \right|^2 \quad \text{(S16)}$$

**Calculation for ghost interference with M-Z interferometer**

Since the M-Z interferometer is integrated into the light path of Signal 2, and Signal 1 side remains unchanged. On Signal 1 side, the quantized field operator at the detection plane is the same as Eq. (S11). On Signal 2 side, an incident horizontal polarization is decomposed into a superposition of left- and right-handed circularly polarized components, the M-Z interferometer makes the two components rotate 180 degrees relative to each other, and then the two components recombine, as a result, the even OAM beam remains the horizontal polarization, while the odd OAM beam attains the vertical polarization [6]. Thus, the even and odd OAM beams can be routed by the following PBS 2, and switched by the HWP in front. So the quantized field operator at the detection plane of Signal 2 is calculated to be a superposition of $E_2^{(+)}(\vec{x}_2)$ in Eq. (S12) and the mirror field of it. The field of the horizontal component can be calculated as,

$$E_{2H}^{(+)} = \frac{1}{2}(E_{s2}^{(+)}(\vec{x}_{s2}) + E_{s2}^{(+)}(-\vec{x}_{s2})) \tag{S17}$$

And the field of the vertical component can be calculated as,

$$E_{2V}^{(+)} = \frac{i}{2}(E_{s2}^{(+)}(\vec{x}_{s2}) - E_{s2}^{(+)}(-\vec{x}_{s2})) \tag{S18}$$

The horizontal and vertical component corresponding to different superposition bases of momentum $k_{s2}$ and $-k_{s2}$. According to the symmetry of field with OAM, the phase at the axisymmetric direction are the same and opposite for even and odd OAM respectively, so $E_{even}^{(+)}(\vec{x}_{s2}) = E_{even}^{(+)}(-\vec{x}_{s2})$ and $E_{odd}^{(+)}(\vec{x}_{s2}) = -E_{odd}^{(+)}(-\vec{x}_{s2})$. There is a prospect that the ghost interference of photon pairs with even and odd OAM will be detected in the horizontal and vertical components respectively.

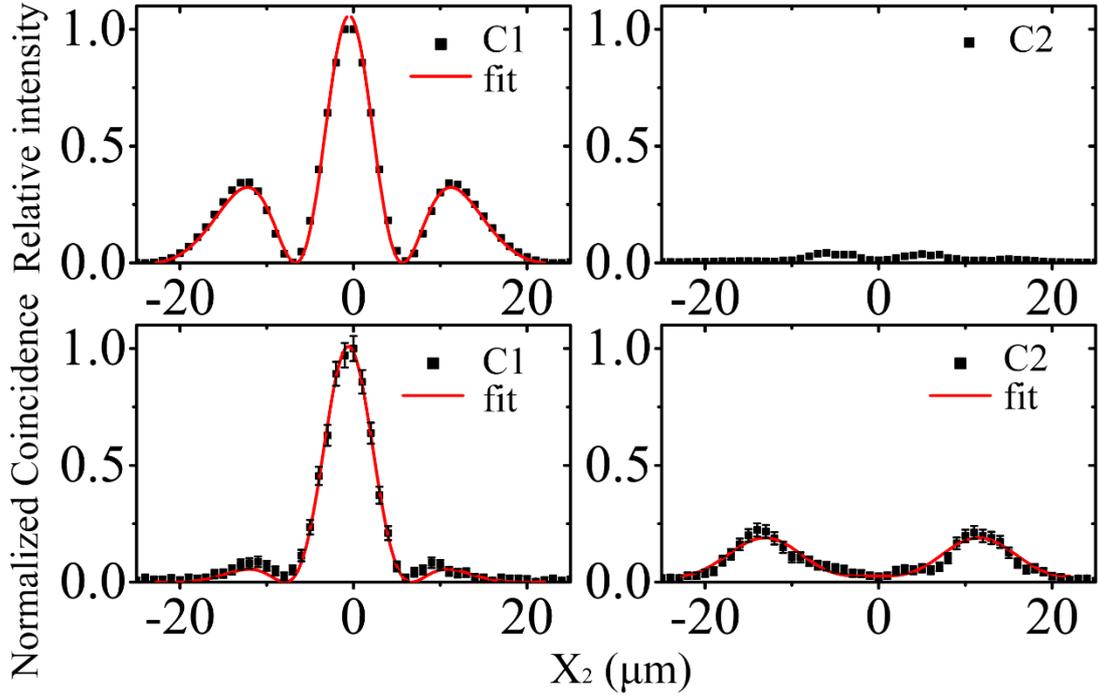

FIG. S1. The upper two figures are the measured interference data of coherent filed from coupler 1, C1 and C2 corresponding to horizontal and vertical components. The red curves are the theoretical fitted results. The lower two figures are the measured ghost interference data of photon pair with 0 OAM. The error bars represent statistical error of ±1 standard deviation.

We measured the interference of coherent field from coupler 1, which is consistent with the theory, the interference is detected only in horizontal components (See the upper two figures in FIG. S1). While for the ghost interference detection (See the lower two figures in FIG. S1), the M-Z interferometer added here led to the change of interference pattern, and some vertical components are detected. The main reason is that for single photon the mirror state corresponding to different position, so the recombined state is different with the initial state, which is a superposition of different position. We realize that the separation of state by the M-Z interferometer is based on its' symmetry, which depends on OAM and linear momentum DOFs. We can calculate the two-photon correlation function for Gaussian by inserting Eq. (S17) and Eq. (S18) into Eq. (S13). The calculation result in,

$$G^{(2)}(x_{s2}) \propto \left| \frac{\sigma_p \sigma_x \omega_0}{2\sqrt{8\pi^2 \omega_0^2 + f^2(\sigma_p^2 + 4\sigma_x^2)\lambda^2}} \exp\left[ -\frac{f^2 x_{s2}^2 (\pi^2(\sigma_p^2 + 4\sigma_x^2)\omega_0^2 + 2f^2 \sigma_p^2 \sigma_x^2 \lambda^2)}{f_A^2 (8\pi^2 \omega_0^2 + f^2(\sigma_p^2 + 4\sigma_x^2)\lambda^2)} \right] \right.$$

$$\times \left( erfc\left[ \frac{-2if^2 \pi(\sigma_p^2 - 4\sigma_x^2)\omega_0^2 \lambda x_{s2} + f_A \omega_b (8\pi^2 \omega_0^2 + f^2(\sigma_p^2 + 4\sigma_x^2)\lambda^2)}{2ff_A \omega_0 \lambda \sqrt{(\sigma_p^2 + 4\sigma_x^2)(8\pi^2 \omega_0^2 + f^2(\sigma_p^2 + 4\sigma_x^2)\lambda^2)}} \right] \right.$$

$$+ erfc\left[ \frac{2if^2 \pi(\sigma_p^2 - 4\sigma_x^2)\omega_0^2 \lambda x_{s2} + f_A \omega_b (8\pi^2 \omega_0^2 + f^2(\sigma_p^2 + 4\sigma_x^2)\lambda^2)}{2ff_A \omega_0 \lambda \sqrt{(\sigma_p^2 + 4\sigma_x^2)(8\pi^2 \omega_0^2 + f^2(\sigma_p^2 + 4\sigma_x^2)\lambda^2)}} \right]$$

$$\left. \left. \pm 2 erfc\left[ \frac{f_A \omega_b (8\pi^2 \omega_0^2 + f^2(\sigma_p^2 + 4\sigma_x^2)\lambda^2)}{2ff_A \omega_0 \lambda \sqrt{(\sigma_p^2 + 4\sigma_x^2)(8\pi^2 \omega_0^2 + f^2(\sigma_p^2 + 4\sigma_x^2)\lambda^2)}} \right] \right) \right|^2$$

(S19)

The $\pm$ in the function corresponding to the horizontal (+) and vertical (-) components respectively.

Assuming that the photocurrent from detector for the horizontal and vertical components of Signal 2 are,

$$I_H \propto \left|E_{2H}^{(+)}\right|^2 = (\left|E_2^{(+)}(x)\right|^2 + \left|E_2^{(+)}(-x)\right|^2 + E_2^{(+)}(x)E_2^{(-)}(-x) + E_2^{(-)}(x)E_2^{(+)}(-x))/4$$

$$I_V \propto \left|E_{2V}^{(+)}\right|^2 = (\left|E_2^{(+)}(x)\right|^2 + \left|E_2^{(+)}(-x)\right|^2 - E_2^{(+)}(x)E_2^{(-)}(-x) - E_2^{(-)}(x)E_2^{(+)}(-x))/4$$

The difference between $E_2^{(+)}(x)$ and $E_2^{(+)}(-x)$ causes the change of the measurement basis, which can be revealed by a two-photon coincidence measurement. We conclude that the generated photon pairs are hyper-entangled in the OAM and position-momentum DOFs, however, all the OAM sorting schemes now change the quantum states of photons [7-10], thus affecting the subsequent demonstration of EPR entanglement. We further measure the ghost interference on the superposition bases of Signal 2. In order to project the Signal 2 photons on the superposition bases vectors of $(|G\rangle - i|L\rangle)/\sqrt{2}$ and $(|G\rangle + |L\rangle)/\sqrt{2}$, we load the superposition phase structures on SLM 2 to convert the superposition bases to Gaussian basis. We obtain the corresponding 8 ghost interferences data, which are shown in FIG. S2.

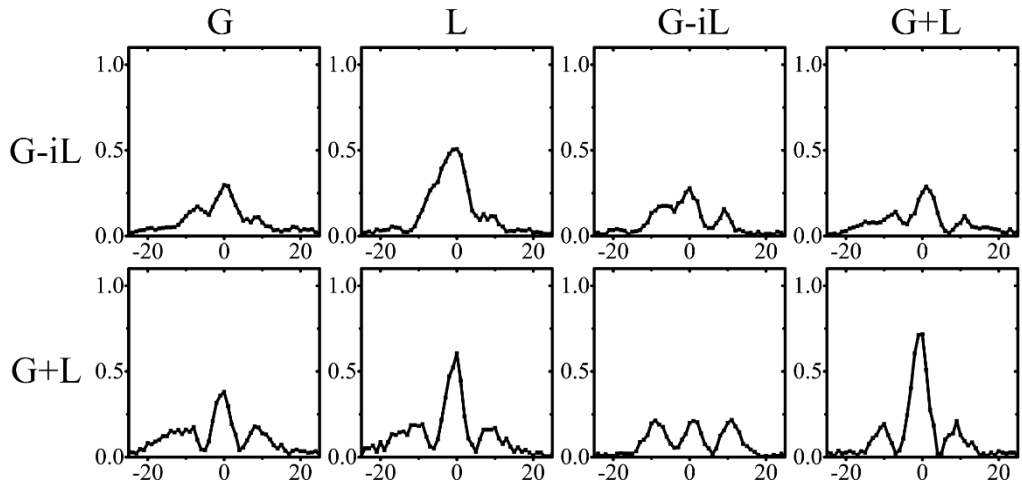

FIG. S2. The ghost interference data on different OAM modes.